# Magnetic order in epitaxial $(La_{2/3}Sr_{1/3}MnO_3)_3/(BaTiO_3)_3$ trilayer heterostructures

E. Rebolini[a], H. Himanshu[a,b,c], K. Beauvois[d], S. Grenier[e], B. Mercey[f], B. Domenges[f,†], B. Ouladdiaf[a], C. Simon[b,*], and M. B. Lepetit[e,a]

[a]*Institut Laue-Langevin, 71 avenue des Martyrs, Grenoble, 38000, France*
[b]*Laboratoire National des Champs Magnetiques Intenses (LNCMI-CNRS, Université Grenoble Alpes), 25 avenue des Martyrs, Grenoble, 38000, France*
[c]*Present address: IBPT-KIT, Karlsruhe Institute of Technology, Karlsruhe, Germany*
[d]*Commissariat à l'energie atomique (CEA), IRIG, MEM, MDN, Univ. Grenoble Alpes, Grenoble, 38000, France*
[e]*Institut Neel, Univ. Grenoble Alpes, CNRS, Grenoble, 38000, France and*
[f]*CRISMAT, ENSICAEN-CNRS UMR6508, 6 bd. Marechal Juin, Caen, 14050, France**
(Dated: October 19, 2026)

The realization of artificial magnetic structures with high epitaxial quality and elevated critical temperatures remains a key challenge for spintronics. Here, we report the fabrication and characterization of epitaxial $(La_{2/3}Sr_{1/3}MnO_3)_3/(BaTiO_3)_3$ trilayer heterostructures grown by layer-by-layer laser ablation. High-resolution transmission electron microscopy, magnetization measurements, and neutron diffraction confirm their exceptional crystalline order and the absence of structural defects. Neutron diffraction reveals no antiferromagnetic at non-zero wavevector Bragg peaks, ruling out most antiferromagnetic configurations with a non zero wave vector Q. Magnetic hysteresis loops instead evidence a ferromagnetic order with a remanent moment of $\sim 0.5\ \mu_B$ per Mn atom at room temperature, persisting either in-plane or out-of-plane depending on the applied field. Our findings highlight neutron diffraction as a powerful tool for elucidating the magnetic structure of complex thin films and suggest that weak interlayer coupling through $BaTiO_3$ leads to ferromagnetism with a high critical temperature ($\sim 600$ K).

## I. INTRODUCTION

The development of spintronics has highlighted significant advantages over other conventional electronics [1]. Consequently, there is considerable interest in engineering artificial magnetic structures with good epitaxial quality and critical temperatures exceeding room temperature, such as manganese-based epitaxial superlattices.

For over 15 years, epitaxial heterostructures composed of stacked monolayers of $La_{2/3}Sr_{1/3}MnO_3$ (LSMO) and $BaTiO_3$ (BTO) have been successfully fabricated, exhibiting highly regular crystal structures with minimal structural defects. In 2010, this system was proposed as a prototypical example of magnetism control via orbital order [2]. Substrate-induced strain, which propagates through the entire film thickness (up to several tens of nanometers), imposes an orbital order in LSMO. This order, stabilized by the periodic insertion of $BaTiO_3$ layers, gives rise to a distinctive magnetic structure with an elevated critical temperature (~ 600 K), significantly higher than that of bulk LSMO (~ 370 K). Such a high critical temperature was indeed observed in samples featuring regular trilayer stacking [2]. Additionally, ferroelectric control of magnetism in similar systems has been reported [Bibes2011], highlighting the strong coupling between structural and magnetic degrees of freedom in these heterostructures.

In LSMO, $Sr^{2+}$ doping at $La^{3+}$ sites results in a mixed-valence state of Mn ions ($Mn^{3+}$ and $Mn^{4+}$). The unpaired fourth electron of $Mn^{3+}$ ions is delocalized via a double-exchange mechanism through oxygen orbitals, leading to a 100% spin-polarized half-metallic state [3]. This phenomenon underpins the colossal magnetoresistance of LSMO—characterized by resistivity changes—and enables its application in spintronic devices [1, 4].

In an octahedral crystal field, the 3d orbitals of $Mn^{3+}$ ions exhibit Jahn-Teller degeneracy lifting between the $3d_{z^2}$ and $3d_{x^2-y^2}$ orbitals. When the manganese octahedra undergo an in-plane expansion and out-of-plane contraction, the $3d_{x^2-y^2}$ orbital is stabilized by the ligand field effect. Indeed, the larger Mn-O delocalisation, induced along the $z$ direction by the out-of-plane contraction, increases the energy of the $3d_{z^2}$ compared to the $3d_{x^2-y^2}$ orbital. In such a configuration, the in-plane magnetic coupling is enhanced. Conversely, out-of-plane expansion of the manganese octahedra favors the occupation of the $3d_{z^2}$ orbital that lower the in-plane magnetic coupling. Assuming that the Curie temperature is related to the in-plane coupling in thin films, the former distortion will enhance the Curie temperature compare to bulk, while the later will reduce it. Such an effect was observed in LSMO-BTO herostructure on an STO substrate [2].

While the hysteresis curves at low temperature were promising, the magnetization measurements aiming at probing the disappearance of magnetic order at the critical temperature were inconclusive. Indeed, the high critical temperature (~ 600 K) led to rapid sample degradation in the paramagnetic phase due to oxygen loss. Additionally, the complexity of the measurements — including magnetization (with substantial substrate signal interference), X-ray diffraction (challenging thin-film ge-ometry), electron microscopy (requiring delicate sample preparation), and neutron diffraction (low signal intensity) — induces significant challenges. These difficulties delayed progress for over a decade. Over the last years, a new neutron diffractometer D10+ was developed at the Institut Laue Langevin within the Endurance program which offer new possibilities to explore thin film magnetism with a higher flux and a better adapted resolution. For this reason, we decided to relaunch this study.

* charles.simon@lncmi.cnrs.fr

## II. SAMPLE PREPARATION

Three nominally identical samples (E1, ref. NG1102; E2, ref. NG1105; E3, ref. NG1118) were investigated. These (3-3) samples consist of LSMO/BTO trilayers, each comprising three layers of LSMO and three layers of BTO, repeated 24 times. The heterostructures were deposited on single-crystalline $SrTiO_3$ (STO) substrates, 001-oriented ($5 \times 5$ mm$^2$, 0.5 mm thick) [5]. The crystallographic axes were defined as follows: the c-axis is perpendicular to the substrate plane, while the a- and b-axes are aligned along the two edges of the square substrate, regardless of the actual lattice parameters. We also prepared samples with two $BaTiO_3$ layers instead of three and two LSMO layers instead of three, named E(2-3) and E(3-2) respectively.

The E3 sample was used for the high resolution electron microscopy studies and then destroyed, while E1 was used for neutron diffraction. E1 and E2 were used for the magnetic measurements to test the reproducibility of the results.

## III. MAGNETIC STRUCTURE MODELS

Bulk LSMO displays a ferromagnetic structure with a Curie temperature of 370 K in single-crystal form. Above this temperature, the material goes to a paramagnetic phase. The crystal structure of bulk LSMO is described by the $R\bar{3}c$ space group, with lattice parameters $a_{\text{LSMO}} = b_{\text{LSMO}} = 5.5$ Å (corresponding to 3.889 Å for the perovskite unit cell) and $c_{\text{LSMO}} = 13.3$ Å [3]. This lattice corresponds to four times the perovskite-structure unit cell, doubled and rotated by 45° in-plane, and doubled out-of-plane, with respect to the simple perovskite mesh. Consequently, the [100] and [010] directions of LSMO align with the equivalent diagonals of the $SrTiO_3$ substrate.

For comparative magnetization measurements, a reference ferromagnetic LSMO sample grown on a silicon substrate with a $SrTiO_3$ buffer layer was used [6]. This LSMO sample exhibits a ferromagnetic structure with an easy axis along the [100] direction of the perovskite structure and a magnetic moment of $\sim 3\ \mu_B$ per manganese atom.

The in-plane lattice parameters of the E1-E3 superlattices are $a_{\text{E}} = b_{\text{E}} = 3.905$ Å, indicating excellent epitaxial growth on STO. Given that the STO parameter is 3.905 Å, all LSMO trilayers are under in-plane expansion compared to the bulk LSMO value (3.889 Å). Along the **c**-axis, the periodicity depends on the number of LSMO and BTO monolayers (one perovskite unit cell), with a measured value of 24.10 ± 0.02 Å for the (3-3) configuration. This periodicity was determined via X-ray diffraction measurements along the $00\ell$ direction. Notably, the films exhibit exceptionally high crystalline quality, as confirmed by both electron microscopy (see Fig. 1) and X-ray diffraction analyses (not shown).

The first task is to estimate the atomic positions and the space group of our samples. For the atomic positions, assuming that the interlayer distances (LSMO-LSMO, BTO-BTO, and LSMO-BTO) remain consistent across the five samples, we derived the following cation-to-cation distances from the **c**-axis repetition periods determined by X-ray diffraction :

- LSMO-LSMO = 3.70± 0.02 Å (La to La)
- LSMO-BTO = 4.35± 0.02 Å (La to Ba)
- BTO-BTO = 4.00± 0.02 Å (Ba to Ba)

To further validate these interlayer distances, we analyzed the cross-sectional high-resolution scanning transmission electron microscopy images, spanning the entire thickness of the sample (see Fig. 1). We measured the distances in these images and constrained the Ba-Ba, La-La, and Ba-La distances to be identical across the different layers. It results the following interlayer distances (defined as the difference in altitude along the **c**-axis for the La/La, La/Ba, and Ba/Ba planes):

- LSMO-LSMO = 3.75± 0.10 Å (La to La)
- LSMO-BTO = 4.20± 0.10 Å (La to Ba)
- BTO-BTO = 4.06± 0.10 Å (Ba to Ba)

These measurements are more direct but less precise than the previous ones. We therefore will use the first set of values to compute the magnetic-structure form factor of the neutron diffraction peaks (see next session).

A key question for analyzing the magnetism in these hetero-structures is the knowledge of their space group. Since the samples have a perfect epitaxy with the STO substrate, and that the $BaTiO_3$ layers share the same in-plane symmetry operations with the substrate, we can reasonably suppose that the hetero-structure space group share these symmetry operations, i.e. belong to P4mm. While the precise atomic positions cannot be fully determined at this stage (due to the large number of atoms involved), the magnetic intensities of the neutron diffraction peaks can still be calculated with reasonable accuracy using the above-defined approximate atomic positions (see Appendix A). The next challenge lies in ensuring that the magnetic symmetry is compatible with all the experimental observations, e.g., the ferromagnetic component in the magnetization measurements (see section V) and the neutron intensities. For this purpose we followed the procedure detailed in Ref. 6 for LSMO.

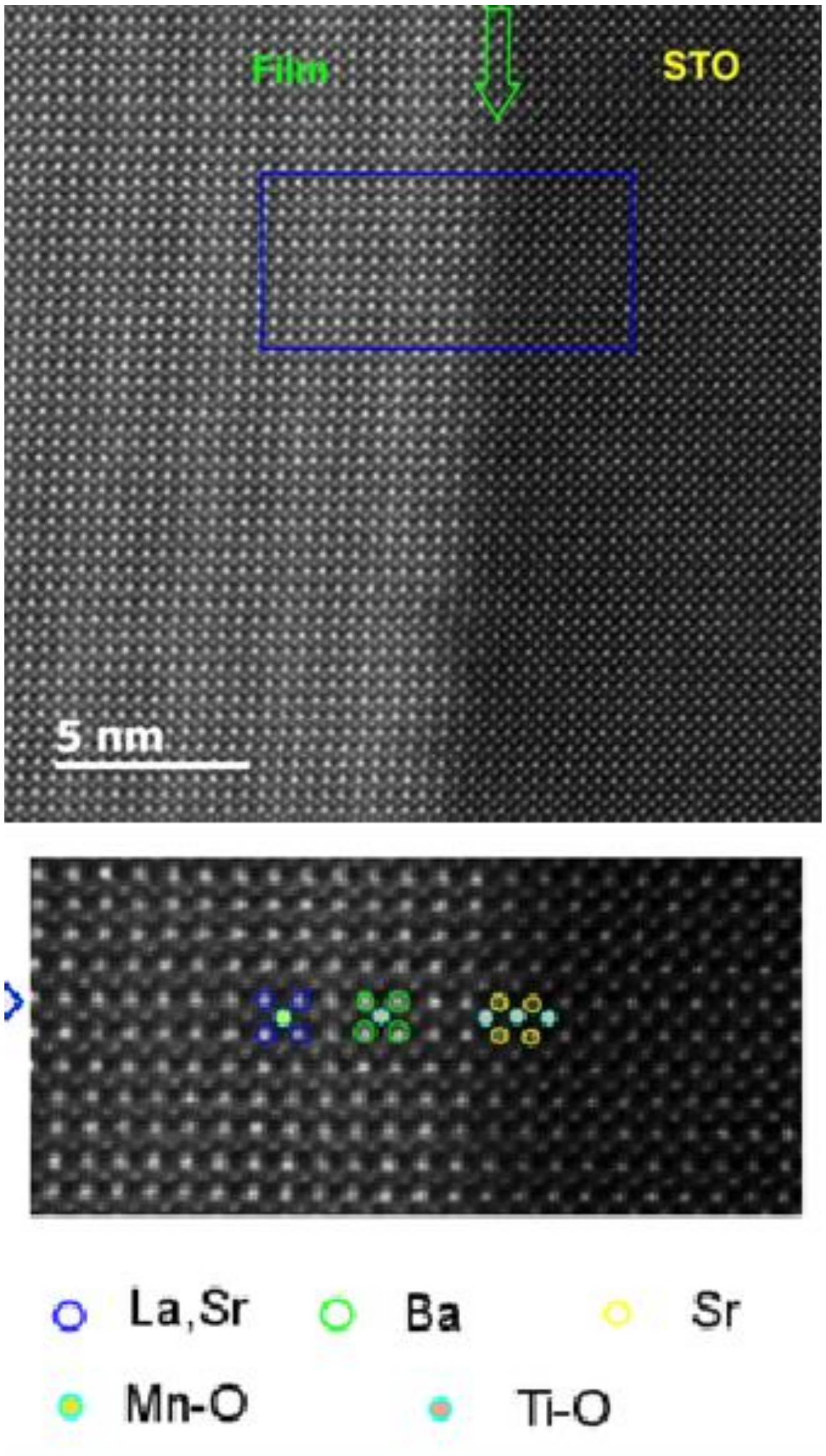

Figure 1: Cross-sectional HAADF-STEM Z-contrast image of the (3,3) sample E3. Upper part: Full view of the heterostructure. Lower part: Enlarged view of the LSMO/STO interface, with atomic species color-coded: blue (La, Sr), green (Ba), cyan (Mn), light blue (Ti). Scale bar: 5 nm.

Group theory tells us that if the Mn atoms belong to a $C_4$ rotation axis along the **c**-direction, and the irreducible representation (irrep) character is +1, the magnetic moment must align along the [001] direction. Conversely, if the irrep character is -1, the magnetic moment must be zero. For Mn atoms not in this symmetrical position, the vector sum of the four magnetic moments will either align along [001] or cancel out, depending on the irrep

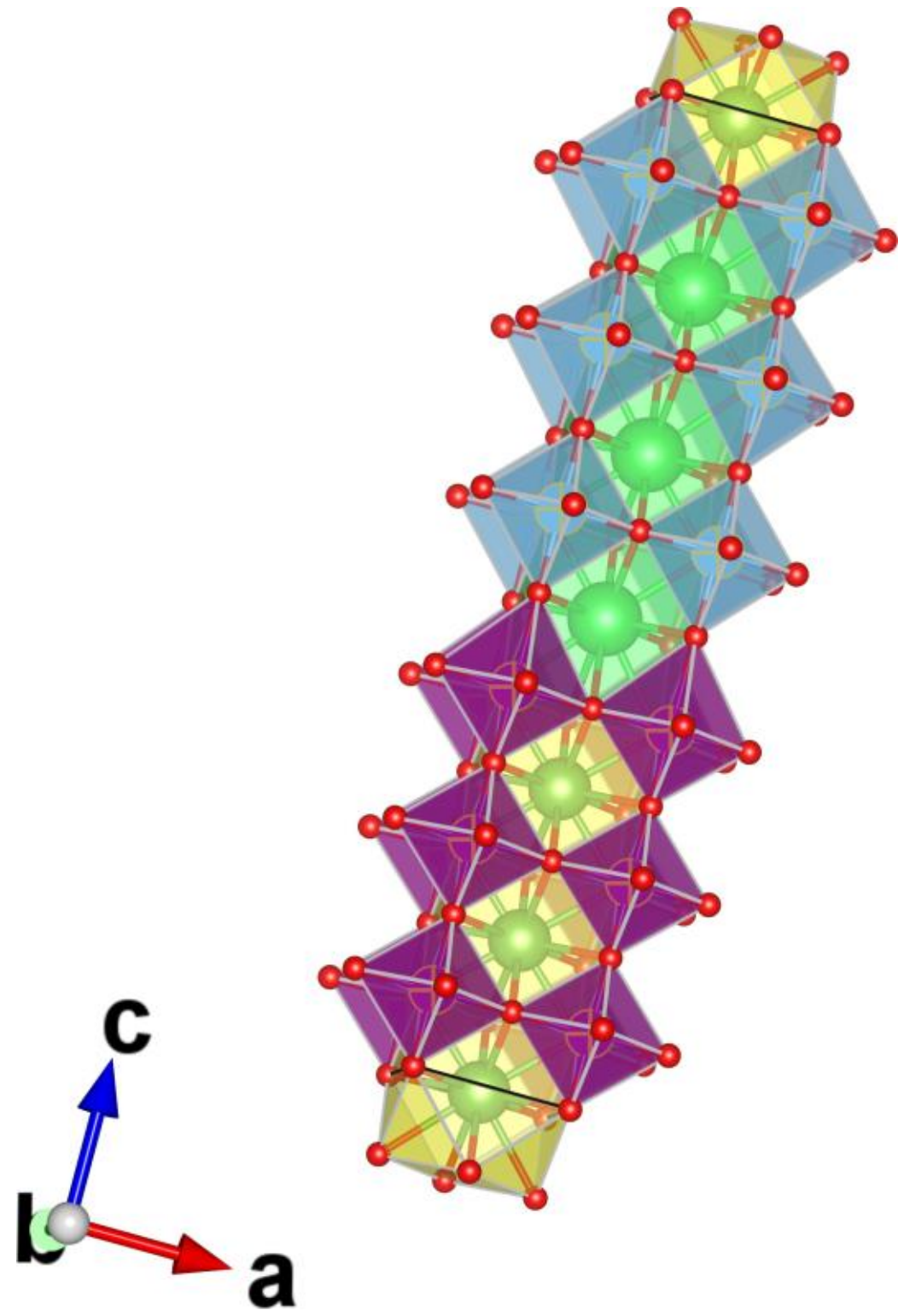

Figure 2: Unit cell of the (3-3) stacking. Oxygen atoms are in red. **$MnO_6$** octahedra (purple) and **$TiO_6$** octahedra (cyan) are shown. La/Sr (yellow) and Ba (Green) atoms are represented as polyhedra. The film is fully epitaxial, with in-plane lattice parameters matching $SrTiO_3$ (3.905 Å).

character.

Based on this analysis, magnetization measurements are essential for selecting the correct space group. The presence of an in-plane ferromagnetic component in the SQUID measurements (see section V) rules out the P4mm space group, as it requires a $C_{4v}$ symmetry axis. Therefore, we must consider subgroups where the 4-fold symmetry is broken. The maximal subgroups of P4mm keeping all symmetry elements except the 4-fold axis are Cmm2 and Pmm2. The existence of a 2-fold rotation axis along the **c**-direction now tells us that the magnetic moment (if the Mn atom lies on the symmetry axis) or the sum of the magnetic moments (if not) should either be along the **c**-direction, $(0, 0, \mu_z)$, or in-plane, $(\mu_x, \mu_y, 0)$ according to the irrep.

From now, we restricted our analysis to easy axes either in-plane ([100]) or along the **c**-direction ([001]). We will therefore set the neutron-diffraction plane of the diffractometer (see section IV) in this (**a**, **c**) sample plane.

Using the Pmm2 group, we calculated the intensities of all purely-magnetic diffraction peaks accessible in our experimental configuration, for all the possible magnetic structure models (see Fig. 3). Due to the in-plane double exchange interaction, we restricted ourselves to models which are ferromagnetically ordered in the (a,b) plane. For this purpose we use the FullProf software, assuming the atomic positions defined in the table of Appendix A. We then selected the most intense peaks for experimental investigation. The calculated intensities are expressed as a percentage of the 0 0 12 peak intensity. Importantly, none of these chosen peaks exhibit ferromagnetic or nuclear intensity (see Table I and Fig. 3). Let us note that the models 3 and 7 displaying a global ferromagnetic alignment between the LSMO/BTO unit cell of the super-lattice, and an antiferromagnetic alignment between the three perovskite unit cells of the LSMO trilayers, cannot be identified by the measurement of these peaks. Indeed, the magnetic intensities of those models are on top of the nuclear peaks, which absolute intensity is not reachable due to sample degradation in the paramagnetic phase.

Table I: Calculated intensities of antiferromagnetic peaks for different magnetic models.

| Model | h k l | Intensities (% of 0 0 12 peak) |
|---|---|---|
| 1 | on top of a nuclear peak | |
| 2 | 0 0 13/2 | 7.52 |
| 3 | on top of a nuclear peak | |
| 4 | 0 0 5/2 | 8 |
| | 0 0 7/2 | 8.2 |
| 5 | on top of a nuclear peak | |
| 6 | 1 0 5/2 | 9.45 |
| | 1 0 7/2 | 9.66 |
| 7 | on top of a nuclear peak | |
| 8 | 1 0 1/2 | 12.1 |

## IV. NEUTRON DIFFRACTION

Neutron diffraction experiments were performed on the D10 and D10+ spectrometers at the Institut Laue-Langevin (D10@ILL & D10+@ILL), in order to decipher between the different magnetic structure models of of the super-lattices. The measurements were done on the E1 sample, at room temperature, following the established methodologies for studying magnetic order in manganites [Chatterji2008]. Data refinement was carried out using the FullProf software [RodriguezCarvajal1993], which allowed us to analyze the magnetic structure factors.

For the diffraction measurements, we utilized both the original D10@ILL spectrometer and its upgraded version, D10+@ILL. The modernization of the monochromator and analyzer allowed us to operate the spectrometer in its high-sensitivity mode, with only a slight reduction of the resolution. This configuration provided an intensity gain of 2.7, as measured on the 0 0 12 peak (comparing D10@ILL and D10+@ILL). The 0 0 12 peak was selected as our intensity reference, with a maximum count of 160 counts/minute on D10+@ILL. In the following, all intensities are expressed as a percentage of the 0 0 12 peak intensity.

The magnetic volume of the sample corresponds to a $5 \times 5\,\mathrm{mm}^2$ area with a thickness of 266 Å (calculated as 24×3×3.70 Å), yielding a total volume of $6.66\times10^{-7}\,\mathrm{cm}^3$. It is essential to remember that the ferromagnetic or ferrimagnetic component cannot be distinguished from nuclear diffraction signals in the D10(+) experiments as the magnetic signal piles up on top of the nuclear one. This limitation arises from the chemical instability of the paramagnetic phase and the absence of polarization analysis on D10(+)@ILL. Indeed, as the compound instability prevents independent measurement of the nuclear peaks in the paramagnetic phase, the intensity of the magnetic component arising on top of these peaks remains inaccessible. We thus limited our study to the search for the existence of purely magnetic peaks, that could whiteness for some antiferromagnetic component, accounting for the reduced total magnetization per atom observed in SQUID measurements (see section V), compared to the expected one in a ferromagnetic phase.

In the D10+ experiment, we focused on the antiferromagnetic peaks of Table I.

Fig. 4 reports the experimental curves measured on D10+ and the associated simulations, calculated using the intensities provided in Table I and taking into account the instrument resolution for the peak width.

- The absence of signal at (0 0 5/2) eliminates the model 4.
- The absence of signal at (0 0 13/2) eliminates the model 2.
- The absence of signal at (1 0 1/2) eliminates the model 8.
- The absence of signal at (1 0 5/2) eliminates the model 6.

We also systematically explored the 0 0 $\ell$ line, but found no evidence of any peaks except the nuclear/ferromagnetic 0 0 6 and 0 0 12 (see Fig. 5).

One plausible explanation for the absence of antiferromagnetic peaks may be that the magnetic coupling through $BaTiO_3$ is extremely weak, leading to weakly coupled LSMO trilayers. In this scenario, each trilayer could exhibit ferrimagnetic order between the pervovskite unit-cells, while remaining weakly coupled along the c-axis, effectively resulting in quasi-2D ferromagnetism, with reduced total moment. Under these conditions, the 0 0 $\ell$ line would reflect only the magnetic form factor of the manganese ions, with the magnetic intensity distributed along the 0 0 $\ell$ row and modulated solely by the magnetic structure factor. However, due to the thinness of the layers, this intensity would be too weak to be detected experimentally. Ferrimagnetic models, such as models 3 and 7, or ferromagnetic models such as models

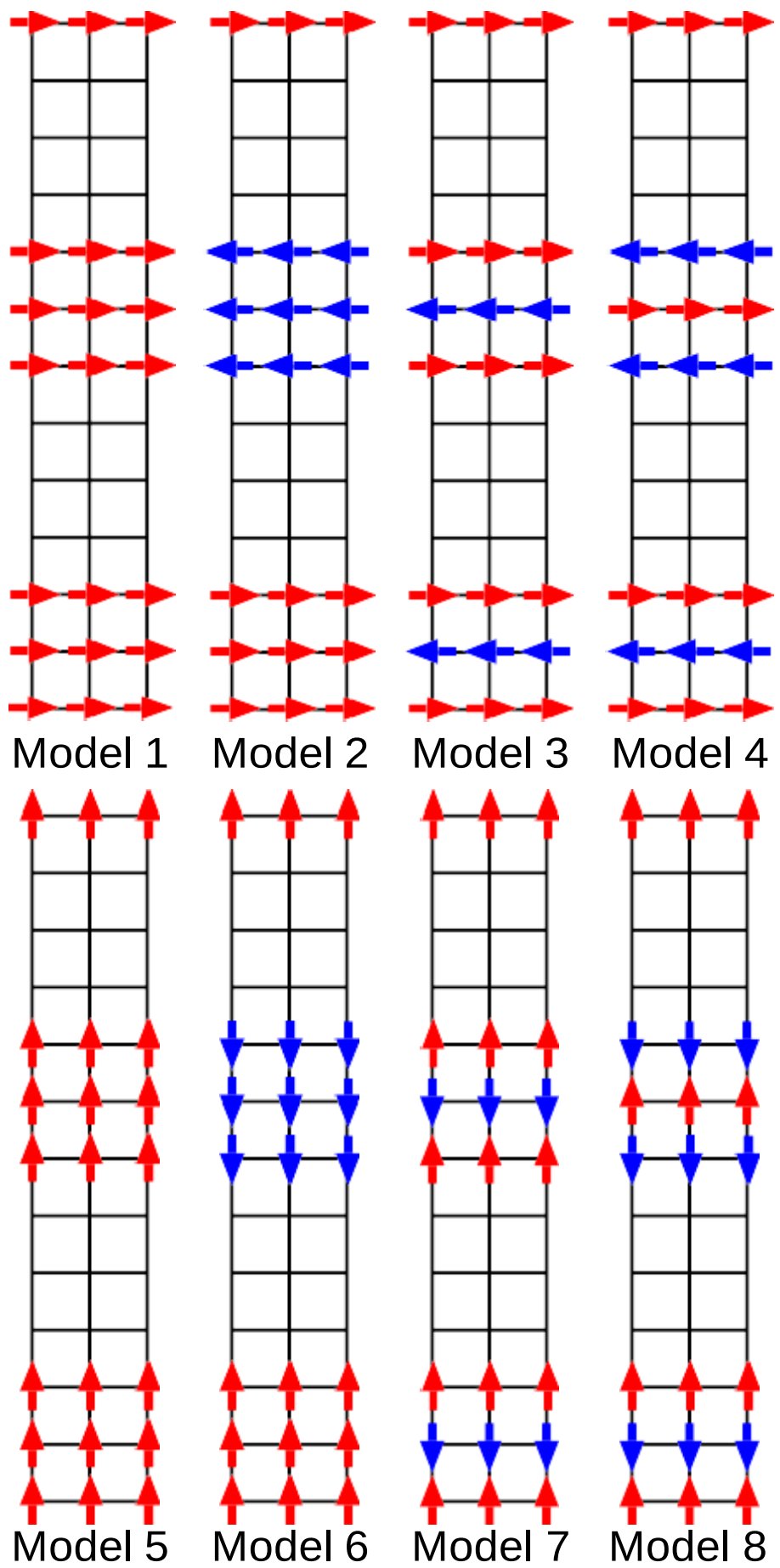

Figure 3: Eight possible magnetic structures of the (3-3) sample compatible with the Pmm2 space group symmetry. The unit cell contains 3 Mn atoms, located along **c**. Red and blue arrows represent spin orientations.

1 and 5 remain possible since there is no predicted intensity at non-integer positions. In these cases, the entire intensity would be added to the existing nuclear peaks.

## V. MAGNETIC MEASUREMENTS

To investigate the existence of a ferromagnetic component in these films, we performed hysteresis cycle measurements, using a SQUID magnetometer (Quantum Design). The samples were first cooled to the target temperature under an applied magnetic field of 5 T. Hysteresis loops were then recorded by cycling the magnetic field along the same direction, from +5 T to -5 T and back to +5 T. Two samples, E1 and E2, were characterized under these conditions, yielding similar results. The substrate signal, measured at the conclusion of the experiments, was systematically subtracted from the data. To do this subtraction, the hysteresis cycles of the substrate were recorded and fitted by an analytical shape (a linear part in field which dominates the signal plus a small squared hysteretic part at low magnetic field) and numerically subtracted to the raw data.

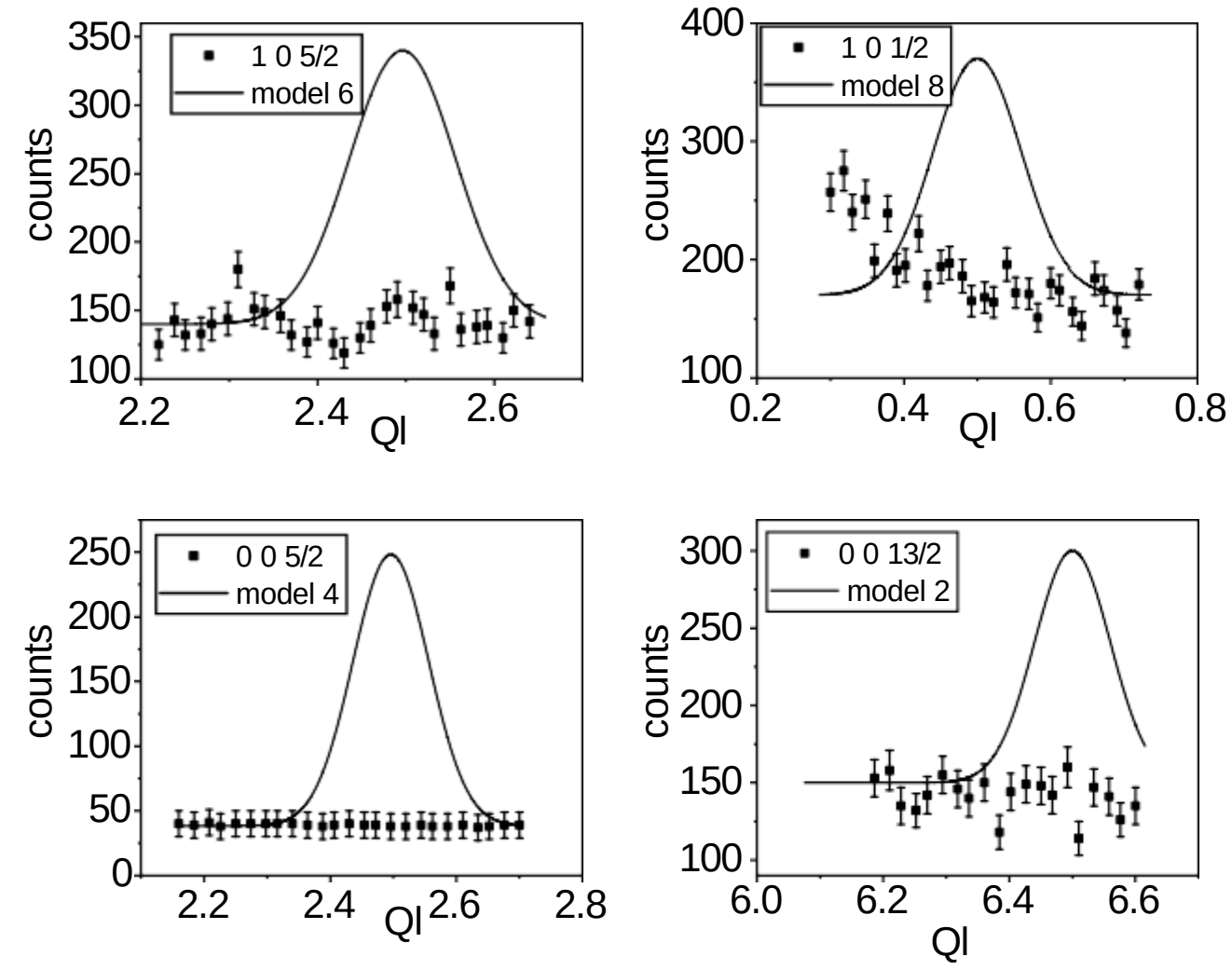

Figure 4: Comparison between the measured counts (points) and the calculated models (solid lines) for the models pictured in Fig. 3. The calculated intensities are normalized to the 0 0 12 peak at 5400 counts (30 minutes counting). Calculations were performed using the FullProf software with the atomic positions provided in Table II of the Appendix A and the intensities provided in Table I.

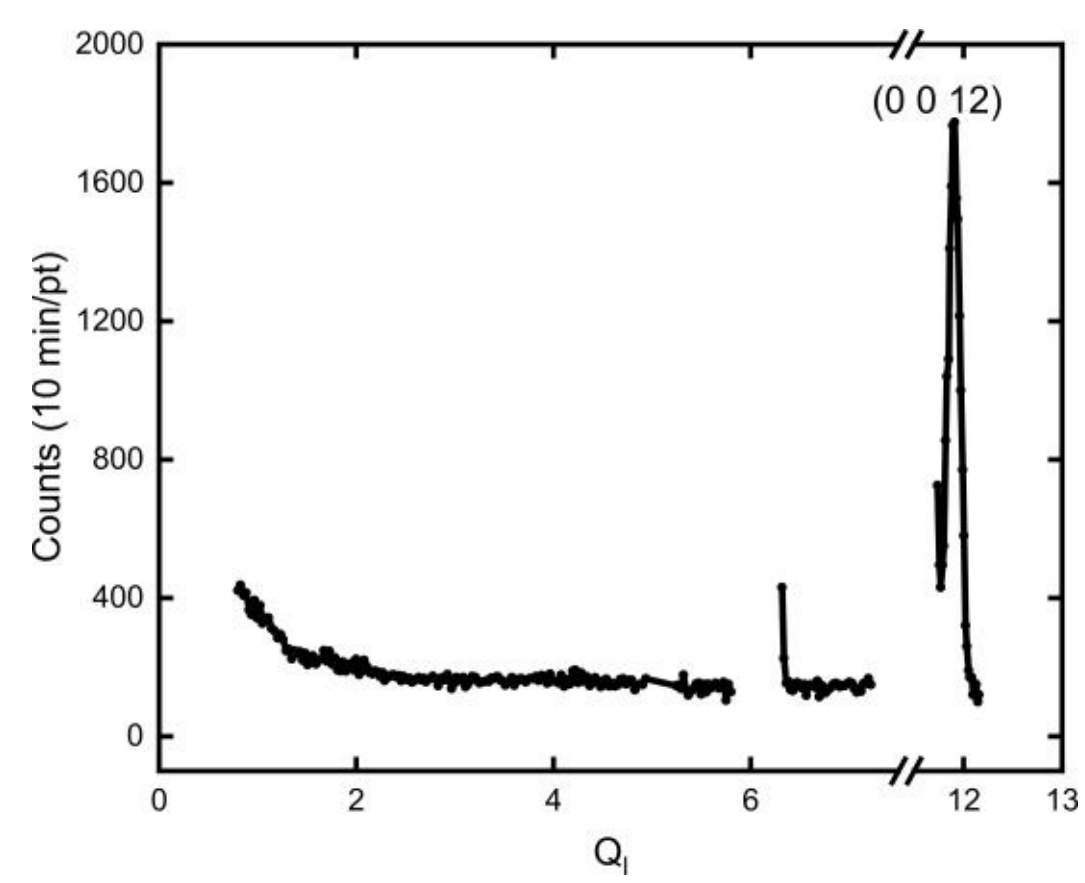

Figure 5: Intensity of the 0 0 $\ell$ line showing the 0 0 12 peak for 10 minutes counting time.

For sample E2, the magnetic response was recorded at 5 K, 10 K, 100 K, and 300 K. At all temperatures, the hysteresis loops exhibit a small ferromagnetic component, characterized by a narrow hysteresis loop that increases progressively toward saturation as the magnetic field is increased (Fig. 6). This behavior is observed with an applied field along the [100] crystallographic direction. The remnant ferromagnetic moment remains small, corresponding to $\sim 0.5\,\mu_B$ per Mn atom. This value (0.65 $\mu_B$/Mn at 5 K) constitutes a minimum bound for the ferromagnetic component, as the existence of magnetic domains can possibly reduce the net magnetization.

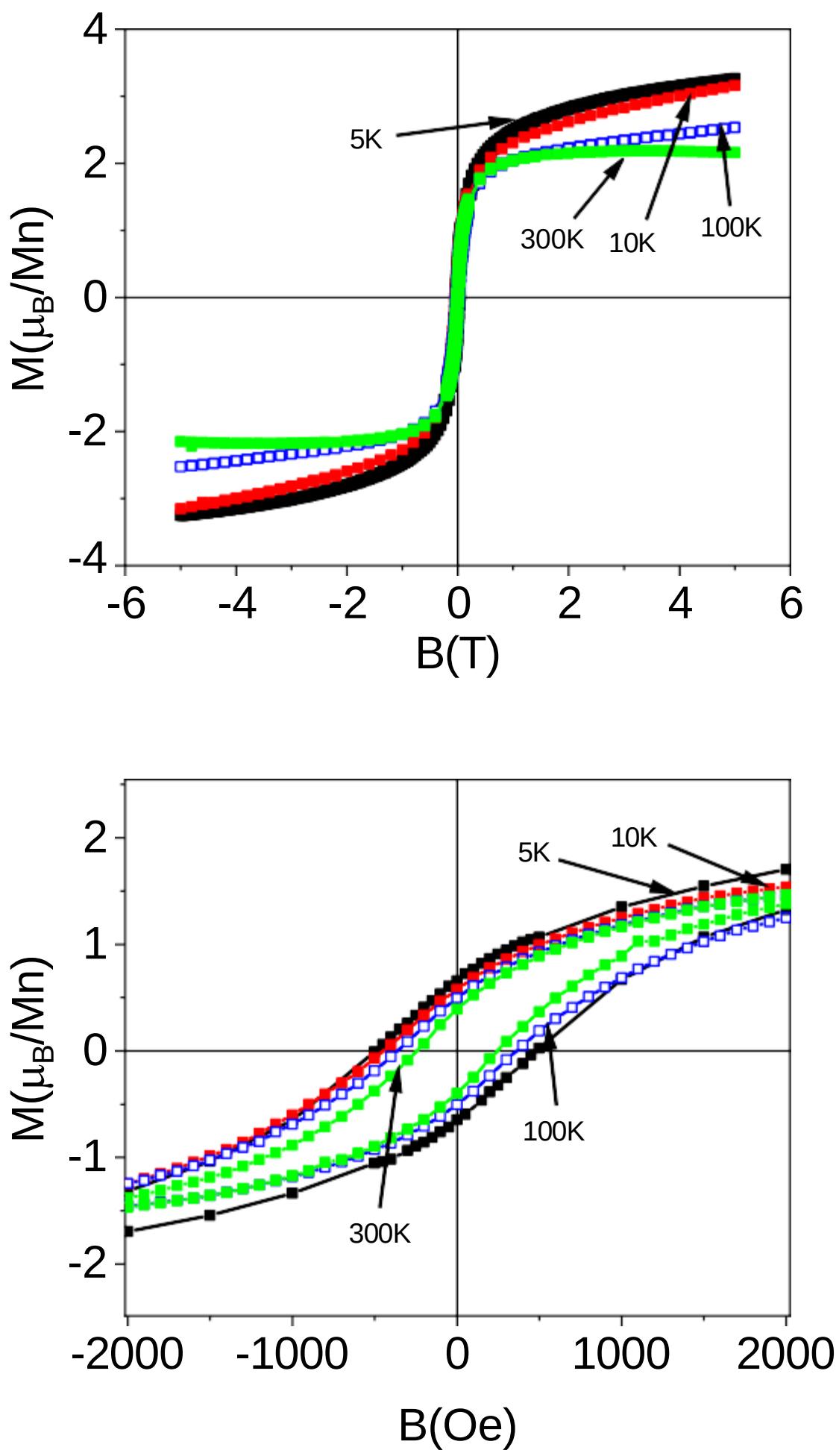

Figure 6: top : Hysteresis cycles of sample E2 (NG1105) with the magnetic field applied along the [100] direction at 5 K, 10 K, 100 K, and 300 K. The signal from the $SrTiO_3$ (STO) substrate was subtracted. bottom : Enlarged view of the low-field region.

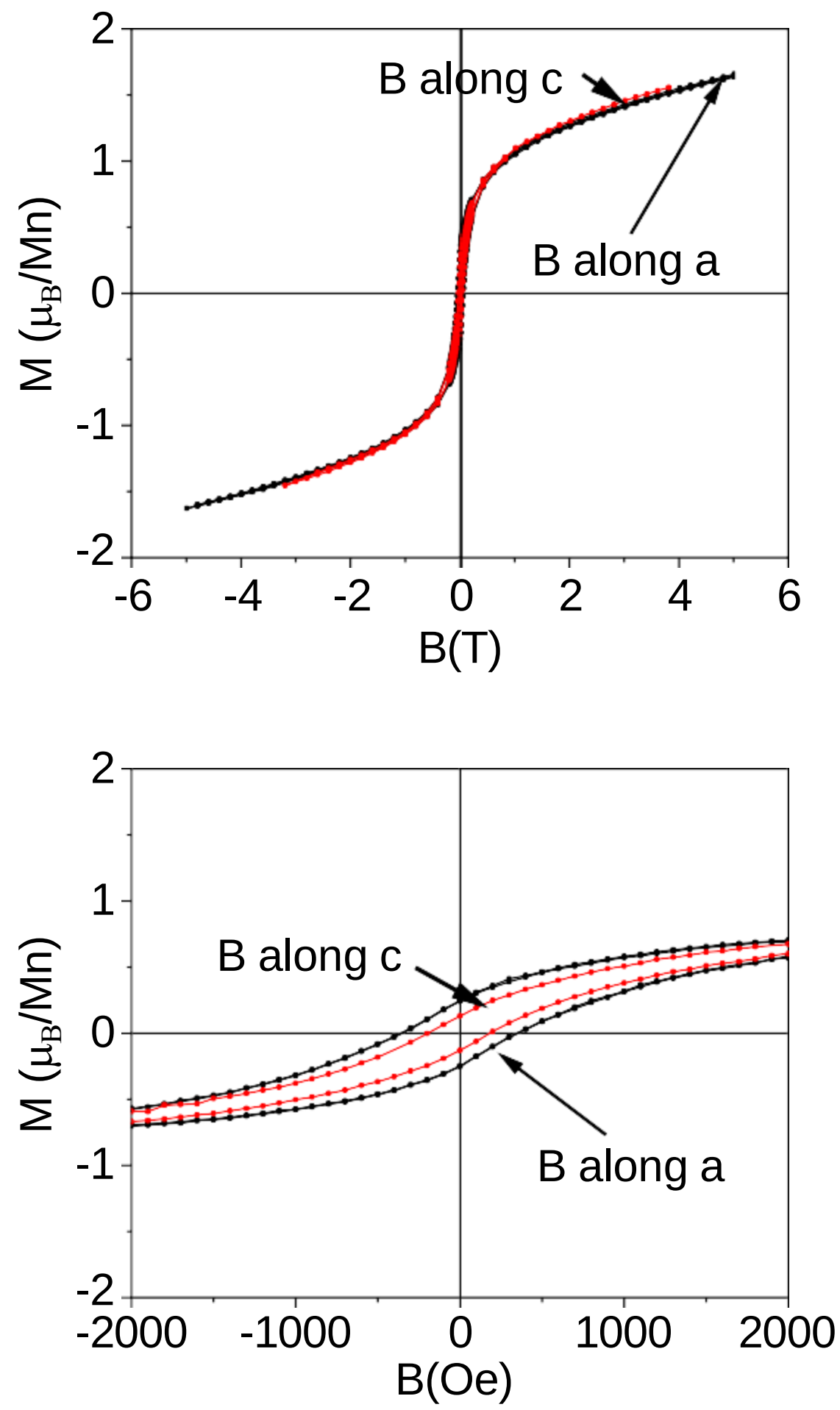

Figure 7: top :Hysteresis cycles compared for sample E1 with the magnetic field applied in the [100] a axis and [001] c axis at 10 K. bottom : Enlarged view of the low-field region. The signal from the STO substrate was subtracted.

Sample E1 exhibits similar results, with a ferromagnetic component smaller than that of E2, probably because the effective surface of the sample is smaller. These two samples are quite similar to the sample studied in Ref. 6.

To determine the easy axis, we compared the three possible orientations of the applied magnetic field (on sample E1) (Fig. 7).

This comparison suggests that the remanent magnetization in zero field depends on the direction in which the magnetic field was applied. When the field was applied along [100], the remanent magnetization is along [100], and when the field was applied along [001], the remanent magnetization is along [001]. Following the group theory already presented, the sample preparation by applying a magnetic field selects different irrep, according to whether it is along the a or c direction. The fact that the remanent, zero-field magnetization remains in the same direction confirms this prediction. This confirms the existence of a two-fold axis in the symmetry group.

Quantitatively understanding the observed hysteresis cycles would require to simulate them for a given model. This is however not straightforward as the moments involve a mixture of $Mn^{3+}$ and $Mn^{4+}$ ions, and one must account for both the Zeeman effect and spin-orbit coupling effects (which cannot be easily approximated by a zero-field splitting term in this case). The magnetic coupling of manganese atoms across the $BaTiO_3$ trilayers is expected to be very small. As a consequence, the demagnetization effects should play an important role in selecting the magnetization orientation in the absence of applied field, and thus ground state.

## VI. SUMMARY AND DISCUSSION OF THE RESULTS

This work studies high-quality epitaxial (3-3) LSMO/BTO heterostructures fabricated via layer-by-layer laser ablation. Combining X-rays scattering and TEM for structural characterizations, group theory, magnetization measurements and neutron diffraction for magnetic characterizations, we were able to propose space group and possible magnetic structures for this highly complex systems.

Neutron diffraction experiments reveals the absence of any antiferromagnetic Bragg peaks with a non zero Q vector. SQUID magnetization measurements confirmed the existence of ferromagnetic components. Associated with group theory, they show the existence in this system of a two fold rotation axis along the c direction.

As a result, the system can be prepared in a specific ground state by an applied field, according to the orientation of the latter.

The observed moment is however reduced compared the expected one in a fully ferromagnetic state. Whether this is due to the sample quality including the existence of magnetic domains, or to a ferrimagnetic ordering within the LSMO trilayers could not be deciphered by the present study. The shape of the hysteresis loops does not suggest the existence of magnetic domains.

The weak magnetic coupling between the trilayers, mediated by $BaTiO_3$, gives rise to ferromagnetic order. This ferromagnetic state persists up to very high temperatures (~600 K), showing the interest of such compounds for spintronics at room temperature.

This work also demonstrates the feasibility of using neutron diffraction to analyze the magnetic structure of novel epitaxial thin films. This approach opens new avenues for the characterization of advanced materials in spintronics and related fields.

## ACKNOWLEDGMENTS

We acknowledge the Institut Laue-Langevin (ILL) for providing beamtime on the D10 and D10+ spectrometers and a PhD grant for Himanshu. We also thank ILL staff for the technical support and fruitful discussions. This work was supported by LNCMI and Neel Institute from CNRS and Univ. Grenoble-Alpes.

† Deceased before submission.

### Appendix A: Atomic Positions in the Unit Cell

The positions of the 30 atoms in the unit cell are given in Table II. Column z indicates the value used in the calculation of the structure factor, with z at the commensurate value closest to the nearest position in z commensurate with 1/12.

Table II: Atomic positions in the unit cell. The z column shows the value used for structure factor calculations, with commensurate values closest to 1/12.

| Atom Type | x | y | z | commensurate z approx. | |
|---|---|---|---|---|---|
| La1 | 0.5 | 0.5 | 0 | 0 | 0 |
| O1 | 0 | 0 | 0 | 0 | 0 |
| Mn1 | 0 | 0 | 0.0778 | 0.0833 | 1/12 |
| O2 | 0 | 0.5 | 0.0778 | 0.0833 | 1/12 |
| O3 | 0 | 0.5 | 0.0778 | 0.0833 | 1/12 |
| La2 | 0.5 | 0.5 | 0.1556 | 0.1666 | 1/6 |
| O4 | 0 | 0 | 0.1556 | 0.1666 | 1/6 |
| Mn2 | 0 | 0 | 0.2334 | 0.25 | 1/4 |
| O5 | 0 | 0.5 | 0.2334 | 0.25 | 1/4 |
| O6 | 0 | 0.5 | 0.2334 | 0.25 | 1/4 |
| La3 | 0.5 | 0.5 | 0.3112 | 0.333 | 1/3 |
| O7 | 0 | 0 | 0.3112 | 0.333 | 1/3 |
| Mn3 | 0 | 0 | 0.3890 | 0.4166 | 5/12 |
| O8 | 0 | 0.5 | 0.3890 | 0.4166 | 5/12 |
| O9 | 0 | 0.5 | 0.3890 | 0.4166 | 5/12 |
| Ba1 | 0.5 | 0.5 | 0.4854 | 0.5 | 1/2 |
| O10 | 0 | 0 | 0.4854 | 0.5 | 1/2 |
| Ti1 | 0 | 0 | 0.5714 | 0.5833 | 7/12 |
| O11 | 0 | 0.5 | 0.5714 | 0.5833 | 7/12 |
| O12 | 0 | 0.5 | 0.5714 | 0.5833 | 7/12 |
| Ba2 | 0.5 | 0.5 | 0.6538 | 0.6666 | 2/3 |
| O13 | 0 | 0 | 0.6538 | 0.6666 | 2/3 |
| Ti2 | 0 | 0 | 0.7456 | 0.75 | 3/4 |
| O14 | 0 | 0.5 | 0.7456 | 0.75 | 3/4 |
| O15 | 0 | 0.5 | 0.7456 | 0.75 | 3/4 |
| Ba3 | 0.5 | 0.5 | 0.8233 | 0.8333 | 5/6 |
| O16 | 0 | 0 | 0.8233 | 0.8333 | 5/6 |
| Ti3 | 0 | 0 | 0.9140 | 0.9166 | 11/12 |
| O17 | 0 | 0.5 | 0.9140 | 0.9166 | 11/12 |
| O18 | 0 | 0.5 | 0.9140 | 0.9166 | 11/12 |